\documentclass{aastex631}

\newcommand{\Add}[1]{\textcolor{black}{#1}}

\shorttitle{A Long-Duration H$\alpha$ and White-Light Superflare on a Young Solar-Type Star EK Draconis}
\shortauthors{Namekata et al.}

\begin{document}

\title{Discovery of a Long-Duration Superflare on a Young Solar-Type Star EK Draconis with Nearly Similar Time Evolution for H$\alpha$ and White-Light Emissions}

\author[0000-0002-1297-9485]{Kosuke Namekata}
\affiliation{ALMA Project, NAOJ, NINS, Osawa, Mitaka, Tokyo, 181-8588, Japan}
\affiliation{Department of Astronomy, Kyoto University, Sakyo, Kyoto 606-8502, Japan}
\affiliation{Astronomical Observatory, Kyoto University, Sakyo, Kyoto 606-8502, Japan}

\author[0000-0003-0332-0811]{Hiroyuki Maehara}
\affiliation{Okayama Branch Office, Subaru Telescope, NAOJ, NINS, Kamogata, Asakuchi, Okayama 719-0232, Japan}

\author{Satoshi Honda}
\affiliation{Nishi-Harima Astronomical Observatory, Center for Astronomy,University of Hyogo, Sayo, Hyogo 679-5313, Japan}

\author[0000-0002-0412-0849]{Yuta Notsu}
\affiliation{Laboratory for Atmospheric and Space Physics, University of Colorado Boulder, 3665 Discovery Drive, Boulder, CO 80303, USA}
\affiliation{National Solar Observatory, 3665 Discovery Drive, Boulder, CO 80303, USA}
\affiliation{Department of Earth and Planetary Sciences, Tokyo Institute of Technology, 2-12-1 Ookayama, Meguro-ku, Tokyo 152-8551, Japan}

\author{Soshi Okamoto}
\affiliation{Astronomical Observatory, Kyoto University, Sakyo, Kyoto 606-8502, Japan}

\author{Jun Takahashi}
\affiliation{Nishi-Harima Astronomical Observatory, Center for Astronomy,University of Hyogo, Sayo, Hyogo 679-5313, Japan}

\author{Masaki Takayama}
\affiliation{Nishi-Harima Astronomical Observatory, Center for Astronomy,University of Hyogo, Sayo, Hyogo 679-5313, Japan}

\author{Tomohito Ohshima}
\affiliation{Nishi-Harima Astronomical Observatory, Center for Astronomy,University of Hyogo, Sayo, Hyogo 679-5313, Japan}

\author{Tomoki Saito}
\affiliation{Nishi-Harima Astronomical Observatory, Center for Astronomy,University of Hyogo, Sayo, Hyogo 679-5313, Japan}

\author{Noriyuki Katoh}
\affiliation{Graduate School of Human Development and Environment, Kobe University, 3-11 Tsurukabuto, Nada-ku, Kobe 657-8501, Japan}
\affiliation{Nishi-Harima Astronomical Observatory, Center for Astronomy,University of Hyogo, Sayo, Hyogo 679-5313, Japan}

\author{Miyako Tozuka}
\affiliation{Nishi-Harima Astronomical Observatory, Center for Astronomy,University of Hyogo, Sayo, Hyogo 679-5313, Japan}

\author{Katsuhiro L. Murata}
\affiliation{Department of Physics, Tokyo Institute of Technology, 2-12-1 Ookayama, Meguro-ku, Tokyo 152-8551, Japan}

\author{Futa Ogawa}
\affiliation{Department of Physics, Tokyo Institute of Technology, 2-12-1 Ookayama, Meguro-ku, Tokyo 152-8551, Japan}

\author{Masafumi Niwano}
\affiliation{Department of Physics, Tokyo Institute of Technology, 2-12-1 Ookayama, Meguro-ku, Tokyo 152-8551, Japan}

\author{Ryo Adachi}
\affiliation{Department of Physics, Tokyo Institute of Technology, 2-12-1 Ookayama, Meguro-ku, Tokyo 152-8551, Japan}

\author{Motoki Oeda}
\affiliation{Department of Physics, Tokyo Institute of Technology, 2-12-1 Ookayama, Meguro-ku, Tokyo 152-8551, Japan}

\author{Kazuki Shiraishi}
\affiliation{Department of Physics, Tokyo Institute of Technology, 2-12-1 Ookayama, Meguro-ku, Tokyo 152-8551, Japan}

\author{Keisuke Isogai}
\affiliation{Astronomical Observatory, Kyoto University, Sakyo, Kyoto 606-8502, Japan}
\affiliation{Department of Multi-Disciplinary Sciences, Graduate School of Arts and Sciences,The University of Tokyo, Komaba, Meguro-ku, Tokyo 153-0041, Japan}



\author{Daisaku Nogami}
\affiliation{Astronomical Observatory, Kyoto University, Sakyo, Kyoto 606-8502, Japan}

\author{Kazunari Shibata}
\affiliation{Kwasan Observatory, Kyoto University, Yamashina, Kyoto 607-8471, Japan}
\affiliation{School of Science and Engineering, Doshisha University, Kyotanabe, Kyoto 610-0321, Japan.}




\begin{abstract}

Young solar-type stars are known to show frequent ``superflares", which may severely influence the habitable worlds on young planets via intense radiations and coronal mass ejections.
Here we report an optical spectroscopic and photometric observation of a long-duration superflare on the young solar-type star EK Draconis (50-120 Myr age) with the Seimei telescope and \textit{Transiting Exoplanet Survey Satellite} (\textit{TESS}).
The flare energy 2.6$\times$10$^{34}$ erg and white-light flare duration 2.2 hr are much larger than those of the largest solar flares, and this is the largest superflare on a solar-type star ever detected by optical spectroscopy.
The H$\alpha$ emission profile shows no significant line asymmetry, meaning no signature of a filament eruption, unlike the only previous detection of a superflare on this star (Namekata et al. 2021, \textit{Nat.Astron}).
Also, it did not show significant line broadening, indicating that the non-thermal heating at the flare footpoints are not essential or that the footpoints are behind the limb. 
The time evolution and duration of the H$\alpha$ flare are surprisingly almost the same as those of the white-light flare, which is different from general M-dwarf (super-)flares and solar flares. 
This unexpected time evolution may suggest that different radiation mechanisms than general solar flares are predominant, as follows: (1) radiation from (off-limb) flare loops, and (2) re-radiation via radiative backwarming, in both of which the cooling timescales of flare loops could determine the timescales of H$\alpha$ and white light. 



\end{abstract}

\keywords{stars: activity --- stars: flare --- stars: solar-type --- stars: individual (EK Draconis) --- Sun: flares}


\section{Introduction} \label{sec:int}
Solar and stellar flares are explosive phenomena on the surfaces observed from radio to X-rays \citep[see][for review]{2011LRSP....8....6S,2017LRSP...14....2B}.
They are thought to be caused by the conversion of magnetic energy into kinetic and thermal energy via magnetic reconnection.
In the case of solar flares, the coronal mass ejections (CMEs), as well as the X-rays and Extreme Ultraviolet (EUV), have severe impacts on the planetary magnetosphere \citep[see][for review]{2021LRSP...18....4T}.
Therefore, magnetic activities on central stars cannot be ignored when discussing planetary habitability and civilization.

The largest flare energy observed on our Sun is approximately 10$^{32}$ erg \citep[e.g.,][]{2012ApJ...759...71E}, and larger flares called ``superflare" with more than 10$^{33}$ erg has never been reported in the modern solar observations \citep[e.g.,][]{2013A&A...549A..66A}.
In recent years, however, observations of solar-type (G-type main-sequence) stars have provided some insight into whether the present-day Sun can produce superflares, or the young Sun could have experienced superflares.
Vigorous searches for stellar flares over the last 30 yrs revealed not only that rapidly-rotating, young solar-type stars show frequent superflares \cite[age of $\sim$100 Myr;][]{1999ApJ...513L..53A}, but also that slowly-rotating, old solar-type stars show superflares with the low occurrence frequency \citep[age of several Gyr;][]{2012Natur.485..478M,2013ApJS..209....5S,2019ApJ...876...58N,2020arXiv201102117O}.
These discoveries suggest that the young Sun could have produced frequent superflares affecting young Earth's environment, and also suggest a possibility that superflares may occur even on the present-day, moderate Sun.
In these context, superflares on solar-type stars have been paid attention from the solar community \citep[e.g.,][]{2013PASJ...65...49S,2013A&A...549A..66A}, planetary community \citep[e.g.,][]{2020IJAsB..19..136A}, and historical and geophysical communities \citep[e.g.,][]{2012Natur.486..240M,2017ApJ...850L..31H}.
However, the mechanism of radiation and mass ejection from superflares, the information needed to answer the above questions of interest, has been unknown.
They can be unveiled by spectroscopic or multi-wavelength observations, although it is rare for solar-type stars.

\Add{EK Draconis (EK Dra) is a young solar-type star with the age of 50--125 Myr. 
It has Sun-like atmosphere with an effective temperature of 5560--5700 K, radius of 0.94 $R_{\odot}$, and mass of 0.95 $M_{\odot}$ \citep{2017MNRAS.465.2076W,2021MNRAS.502.3343S}.
It is rapidly rotating with period of 2.77 d and exhibits frequent stellar flares, so it is considered as a good target of flare observations \citep{1999ApJ...513L..53A,2015AJ....150....7A,Namekata2020Sci}.}
One UV spectroscopic observation of the decay phase of a superflare on EK Dra has been reported using the \textit{Hubble Space Telescope} \citep{2015AJ....150....7A}, which had been the only example of a spectral line observation for solar-type stars before.
Recently, on another solar-type star HII 345 (G8V), simultaneous observations with \textit{Kepler Space Telescope} and \textit{XMM-Newton} have detected X-ray and white-light emission from superflares \citep{2019A&A...622A.210G}. 



Optical spectroscopic observations are also essential to capture chromospheric phenomena (accompanied by UV radiation) and filament/prominence eruptions (indirect evidence of CMEs) \citep[cf.][for solar and M-dwarf flares]{1984SoPh...93..105I,2013ApJS..207...15K,2020PASJ...72...68N,2020PASJ..tmp..253M}.
Our previous study \citep{Namekata2020Sci} reported the first detection of the optical H$\alpha$ spectra of a superflare of $\sim 10^{33}$ erg on the young solar-type star EK Dra.
Surprisingly, the H$\alpha$ spectra show a blueshifted absorption as evidence of a filament eruption, which has dramatically advanced our understandings.
However, the mechanism of the superflare radiation is still unknown because in the first event by \cite{Namekata2020Sci},  the flare emission was short-lived ($\sim$16 min) and had not been thoroughly investigated.
In this Letter, we report detection of the optical spectra of another gigantic superflare event on EK Dra, which enables us to investigate the radiation mechanism.
We show our observational summary in Section \ref{sec:ana}, results in Section \ref{sec:res}, and discussion in Section \ref{sec:dis}.


\section{Observations and Analysis} \label{sec:ana}


We conducted optical spectroscopic and photometric monitoring observations of the young solar-type star EK Dra from February to April 2020. 
The spectroscopic data were obtained by the 3.8-m Seimei telescope \citep{2020PASJ...72...48K} (see Section \ref{sec:seimei}) and photometric data were obtained by \textit{Transiting Exoplanet Survey Satellite} \citep[\textit{TESS};][]{2015JATIS...1a4003R} (see Section \ref{sec:tess}). 
Through this campaign, we succeeded in obtaining optical spectroscopic data of two superflares on a solar-type star, simultaneously with \textit{TESS} photometry.
One event on April 5, 2020 was already reported by \citet{Namekata2020Sci}, and another event on March 14, 2020 will be reported in this Letter.


\subsection{TESS} \label{sec:tess}

\textit{TESS} observed EK Dra (TIC 159613900) in its sector 14-16 (18 July 2019-6 October 2020) and 21-23 (21 January 2020-15 April 2020).
EK Dra was observed with the 2-min time cadence during this period \citep{2015JATIS...1a4003R,Fausnaugh2020}. 
The \textit{TESS} light curves are shown in Figure \ref{fig:1}.

We performed the automatic flare detection as follows \citep[see,][]{2020PASJ..tmp..253M}.
We analyzed the TESS Pre-search Data Conditioned Simple Aperture Photometry (PDC-SAP) light curves retrieved from the MAST Portal site (\url{ https://mast.stsci.edu/portal/Mashup/Clients/Mast/Portal.html }).
We first removed background rotational brightness variations using the Fast Fourier Transformation with a low-pass filter.
We used the cut-off frequency of 3.0 d$^{-1}$ in the first process.
As the flare candidate, we selected the data points with the following criteria:
(1) the peak residual brightness of the data point is higher than five times the \textit{TESS} photometric errors, and
(2) at least two consecutive data points exceed the three times the \textit{TESS} photometric errors.
A complex flare having multiple peaks was manually recognized as a single flare.
Also, for flares of particular interest, we manually modeled the background component after removing the flare periods and changed the cutting frequency in low pass filter (for Figure \ref{fig:2} (A), we use the cutting frequency of 10 d$^{-1}$).

In addition, we analyzed the \textit{TESS} pixel-level data. 
No centroid motions are found during the superflares, which suggests it is associated with the stellar system rather than an instrumental systematic error or contamination from scattered background light or a distant star.
Following the method of  \cite{2013ApJS..209....5S}, the white-light flare's bolometric energy is derived by assuming the flaring spectra of 10000 K blackbody radiation and \textit{TESS} response function \citep{2015JATIS...1a4003R}.

\subsection{3.8-m Seimei telescope} \label{sec:seimei}

We introduce the utilization of low-resolution spectroscopic data from KOOLS-IFU \citep[Kyoto Okayama Optical Low-dispersion Spectrograph with optical-fiber Integral Field Unit;][]{2019PASJ...71..102M} installed at the 3.8-m Seimei Telescope \citep{2020PASJ...72...48K} at Okayama Observatory of Kyoto University. 
KOOLS-IFU is an optical spectrograph with a spectral resolution of R $\sim$ 2,000 covering a wavelength range from 5800-8000 {\AA}, including the H$\alpha$ line (6562.8 {\AA}).

The observation was conducted between February to April, 2020 (\textit{TESS} Sector 21-23), and the observational periods are indicated with blue color in Figure \ref{fig:1} (C).
The exposure time was set to be 30 sec for these nights. 
The data reduction follows the prescription in \cite{2020PASJ...72...68N,Namekata2020Sci} with \textsf{IRAF} and \textsf{PyRAF} packages.
Only the data of particular interest are shown in Figure \ref{fig:2}.
The original flare and template pre-flare H$\alpha$ spectra are shown in Figure \ref{fig:3}.
We measure the H$\alpha$ equivalent width (hereafter ``EW", which is a H$\alpha$ emission integrated for 6562.8--10 {\AA} $\sim$ 6562.8+10 {\AA} after being normalized by nearby continuum level) and plotted the light curve in Figure \ref{fig:2}. 
For the flare data, the H$\alpha$ radiated energy is calculated by multiplying the enhanced H$\alpha$ EW by the continuum flux and integrating in time.
The continuum flux of EK Dra around H$\alpha$ is derived as 1.57 W m$^{-2}$ nm$^{-1}$ at 1 AU with the stellar distance given by Gaia Data Release 2 \citep{2018A&A...616A...2L}.



\section{Results} \label{sec:res}


\subsection{Statistical properties of white-light flares on EK Dra detected by \textit{TESS}}\label{res:stat}

\begin{figure}
\epsscale{1}
\plotone{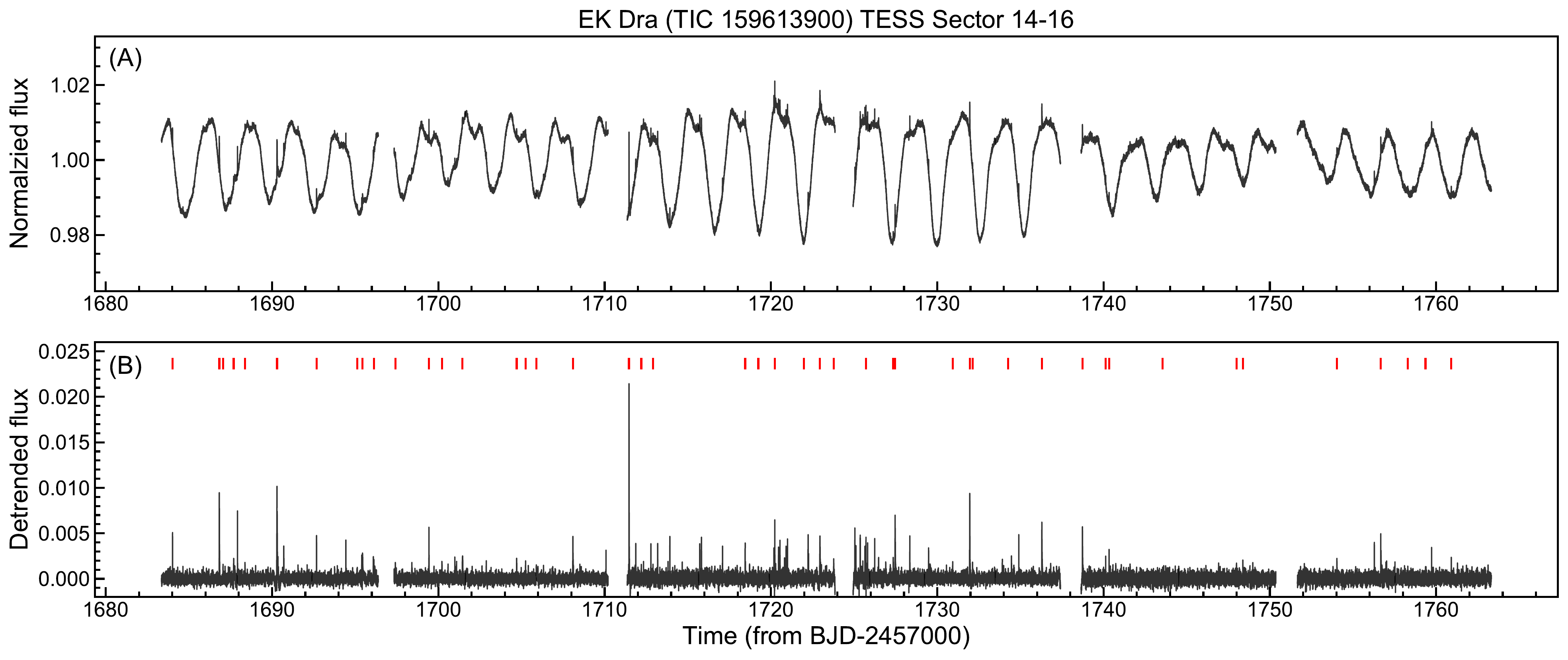}
\plotone{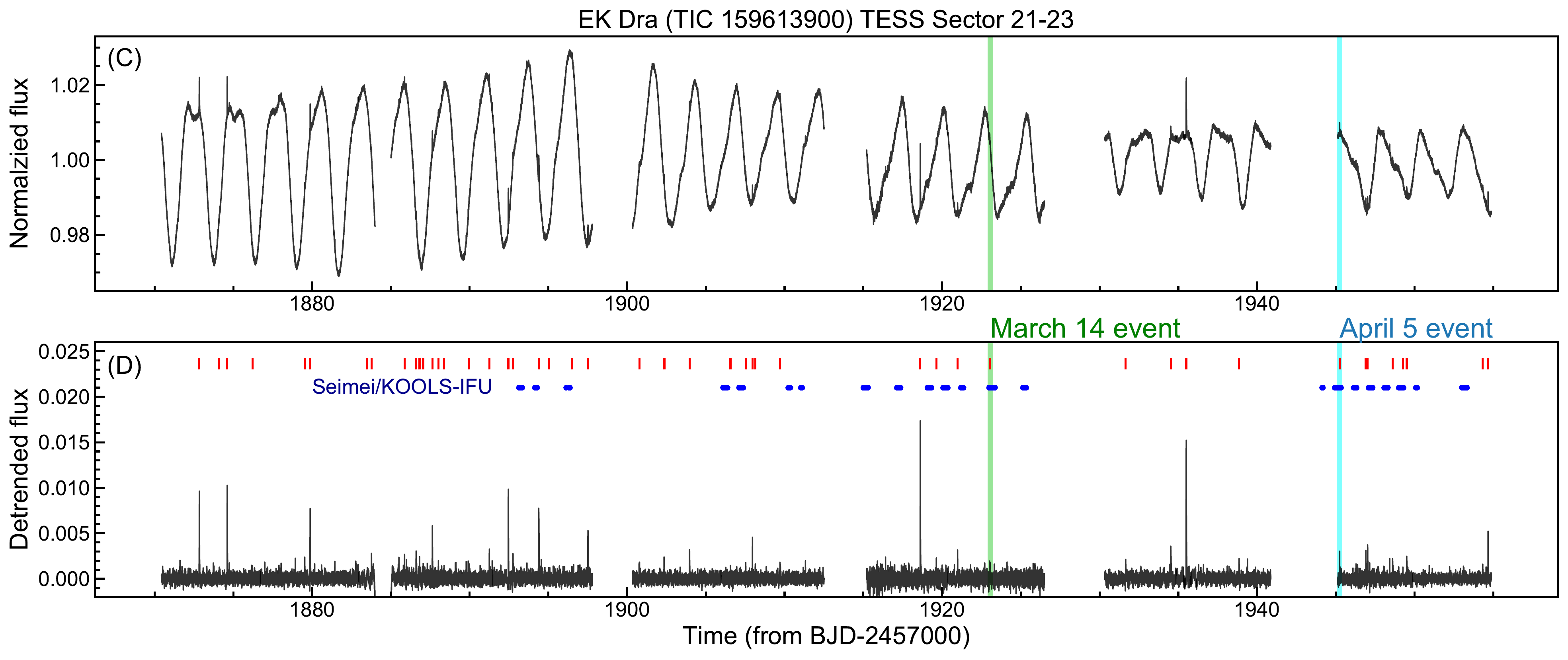}
\epsscale{0.5}
\plotone{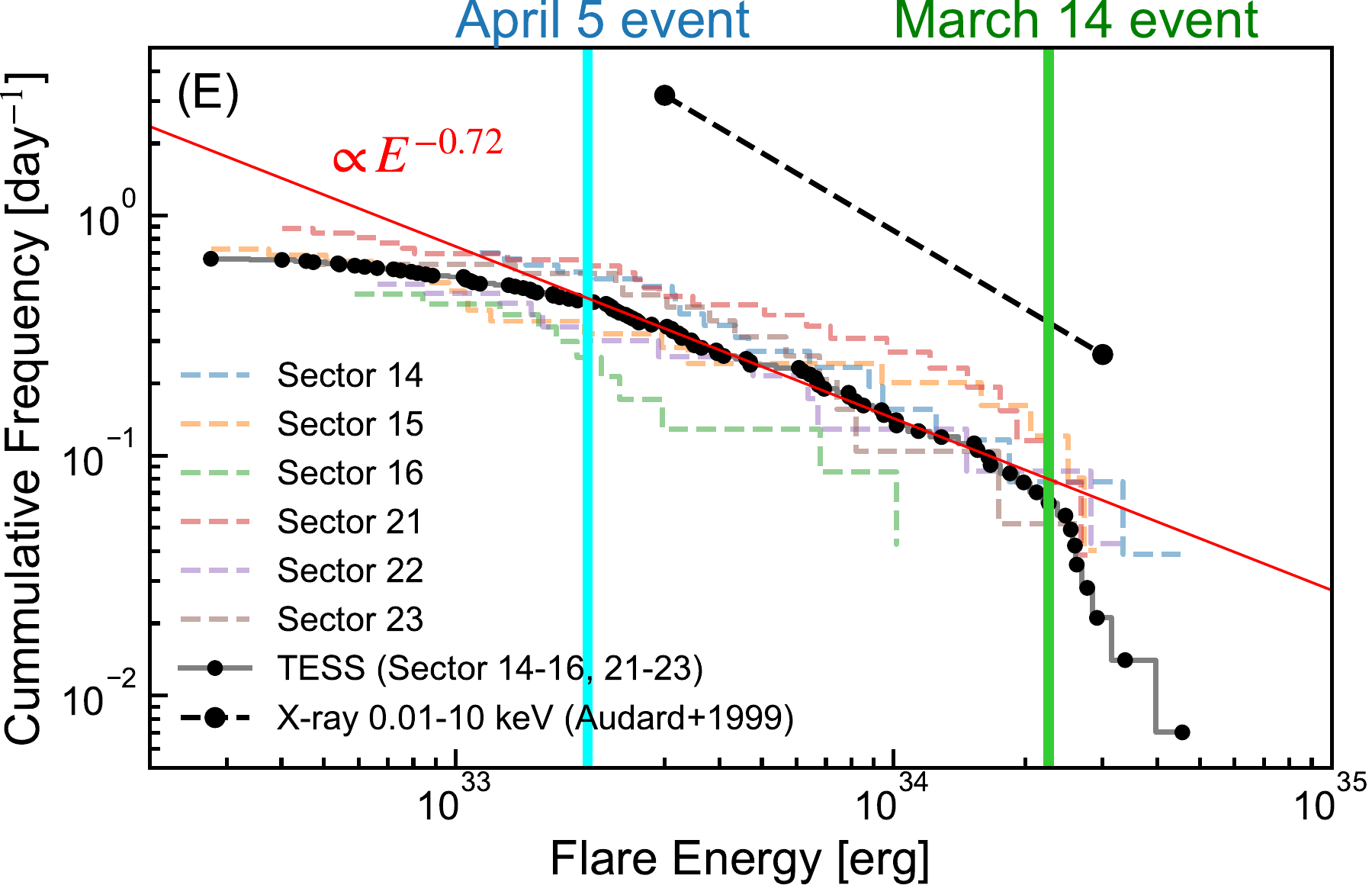}
\plotone{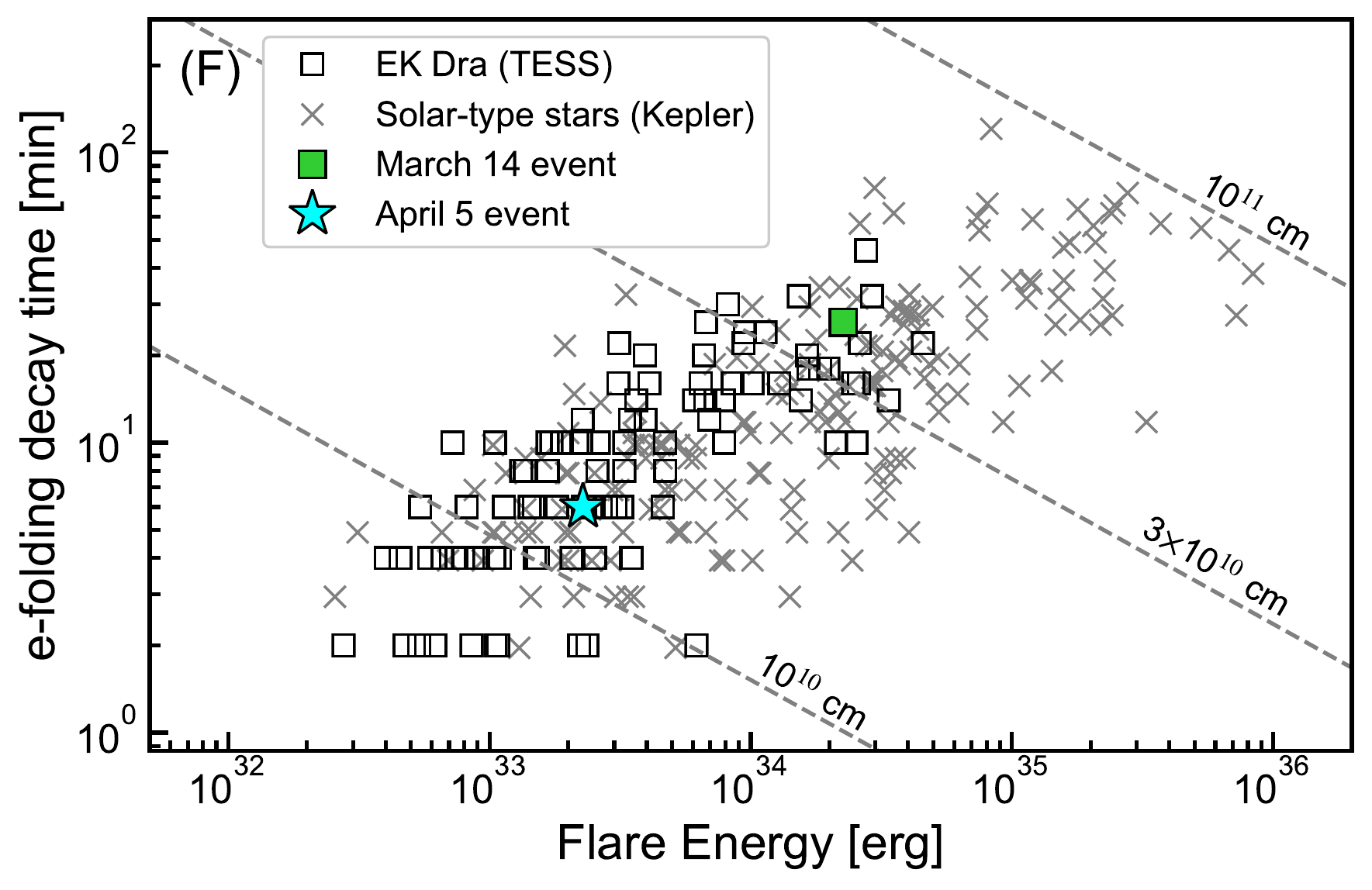}
\caption{Superflares on EK Dra detected by \textit{TESS} (Sector 14-16 and Sector 21-23). 
(A) The \textit{TESS} light curve of Sector 14-16 (18 July 2019 to 7 October 2019) normalized by the average flux. (B) The detrended light curve of the panel (A). The times of the automatically detected flares are indicated by the red lines.
(C) The \textit{TESS} light curve during Sector 21-23 (21 January 2020 to 6 April 2020) normalized by the average flux. (D) The detrended light curve of the panel (C). The periods of the monitoring observations by Seimei telescope/KOOLS-IFU are marked in blue. Flares detected by ground-based telescopes on 14 March 2020 (This Letter) and 5 April 2020 \citep{Namekata2020Sci} are also shown in green and cyan, respectively.
Note that the detrended light curve is automatically obtained by low pass filter without removing the flares, and background subtraction is different from Figure \ref{fig:2} (A) (see, Section \ref{sec:tess} for the analysis).
(E) The flare frequency on EK Dra detected by \textit{TESS}. The flare frequency of EK Dra in X-rays, as determined by \cite{1999ApJ...513L..53A}, is shown as a black line as a reference.
\textcolor{black}{The red solid line corresponds to the fitted one (N($>E$)$\propto E^{-0.72\pm0.01}$) for 2$\times 10^{33}$ to 2$\times 10^{34}$ erg }
(F) The relationship between flare energy and duration as determined by TESS. The gray cross points are the relationship between energy and duration obtained by Kepler \citep{2015EP&S...67...59M}. The theoretical scaling relation with constant flare loop length (10$^{10}$, 3$\times$10$^{10}$, and 10$^{11}$ cm) is shown by the dashed line \citep{2017ApJ...851...91N}.
}
\label{fig:1}
\end{figure}

Figure \ref{fig:1} (A) and (C) show the \textit{TESS} light curves observed in Sectors 14-16 and 21-23, respectively. 
Figure \ref{fig:1} (B) and (D) show the detrended light curves for panels (A) and (C), respectively. 
We detected 94 flares on EK Dra in total, and the automatically detected flares are shown in red.
Figure \ref{fig:1} (E) shows the flare occurrence frequency as a function of flare energy. 
\Add{The relation can be fitted as $N$($\ge E$)$\propto E^{-0.72\pm0.01}$ for $2\times10^{33}$ to $2\times10^{34}$ erg.
Note that the fitted energy range is limited because higher energy range could be affected by the lack of the statistics and energy cutoff, and the lower energy range could be affected by the flare detection sensitivity \citep{2019ApJ...880..105A,2020PASJ..tmp..253M,2020arXiv201102117O}.}
The occurrence frequency of superflares with the energy of $>$ 10$^{33}$ erg is 0.56 events per day, and that with the energy of $>$ 10$^{34}$ erg is 0.14 events per day.
The flare occurrence frequency varies for each sector (Figure \ref{fig:1} (E)), and \Add{it is reported that the variation of flare frequency is positively correlated with the spot filling factor (Supplementary Fig. 2 in \cite{Namekata2020Sci}).}
Also, note that the X-ray flare frequency of EK Dra of $\sim$4.2 flares d$^{-1}$ at 1995 \citep[a dashed line in Figure \ref{fig:1} (E)][]{1999ApJ...513L..53A} is much larger than that white-light flare frequency at 2019-2020.
\Add{This can be a cause by the long-term activity changes \citep[e.g., 8.9 yr activity cycle is reported by][]{2018A&A...620A.162J}, or a difference in energy partition for X-ray and white-light emission \citep[e.g.,][]{2011A&A...530A..84K,2012ApJ...759...71E}.}


Figure \ref{fig:1} (F) shows the relationship between flare energy $E_{\rm flare}$ and decay time $\tau_{\rm decay}$ on EK Dra observed by \textit{TESS} ($\tau_{\rm decay} \propto (E_{\rm flare})^{0.49\pm 0.04}$) and that on the solar-type stars observed by \textit{Kepler} \citep[with 1 min cadence;][]{2015EP&S...67...59M}.
We found that the timescales of the EK Dra flares are comparable to those of the superflares in many solar-type stars discovered by \textit{Kepler}.

\subsection{The gigantic H$\alpha$ and white-light superflare on March 14, 2020}\label{res:flare}

\begin{figure}
\epsscale{1.2}
\plotone{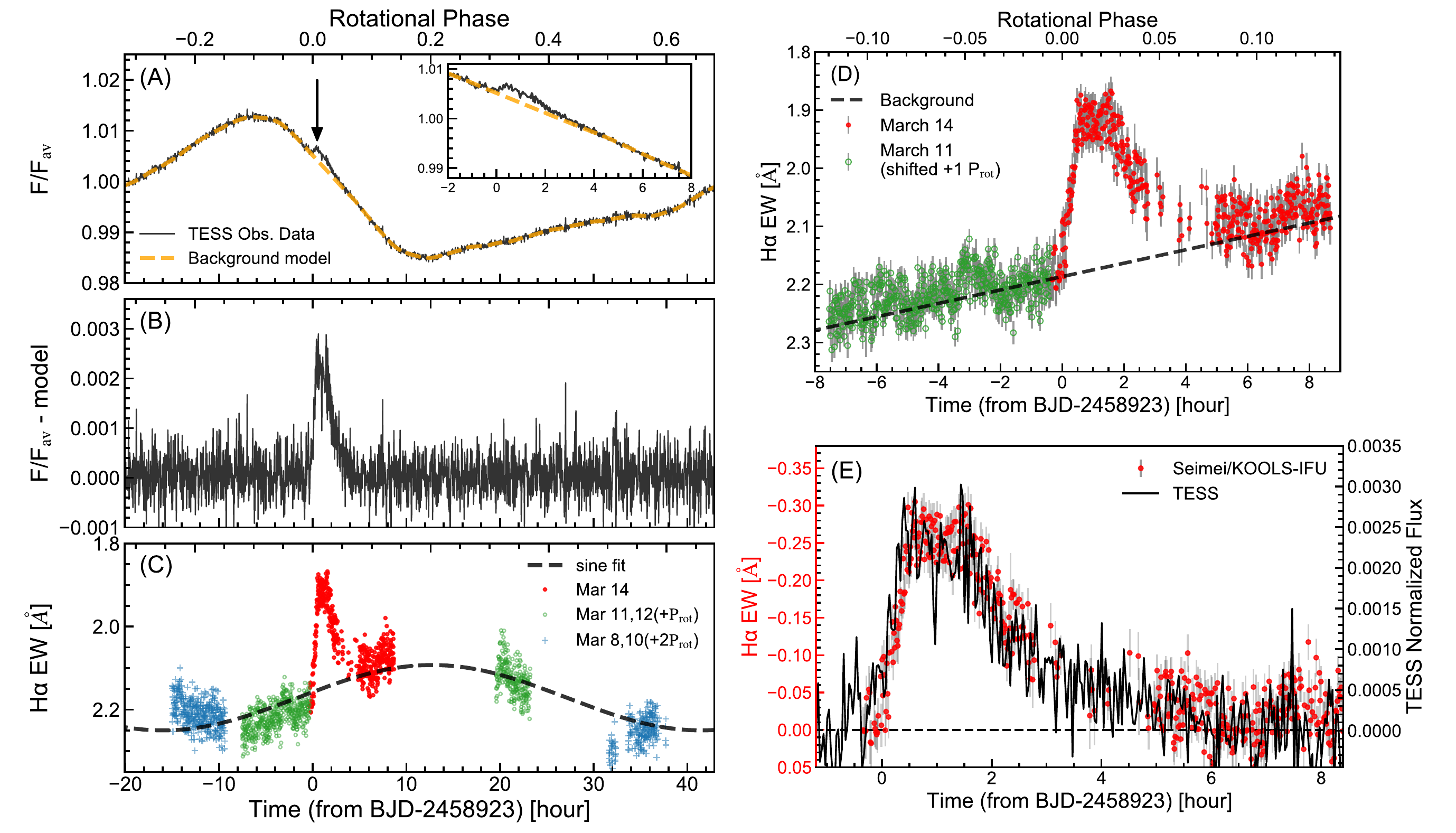}
\caption{White light and H$\alpha$ light curves of the superflare and rotation variations on EK Dra observed on March 14, 2020.
(A) The black line is the observed \textit{TESS} light curve at the rotation phase when the superflare occurred. The orange dashed line is one fitted with the FFT and low-pass filter. The upper right in panel (A) is an enlarged light curve where the flare occurred.
(B) The detrended \textit{TESS} light curve (i.e., the black minus orange line in panel (A)).
(C) The rotational modulation of the H$\alpha$ EW folded by the rotational period ($P_{\rm rot}$ = 2.62 d and t$_0$ = BJD-2458923). The different colors plot different rotational phases, and the red data are the rotation phase when the superflare occurred. The dashed line is a line fitted with a simple sine function for these rotation variations.
The EW is the integrated H$\alpha$ absorption between 6552.8 {\AA} and 6572.8 {\AA}, normalized by the continuum level of 6517.8-6537.8 {\AA} and 6587.8-6607.8 {\AA}.
(D) The H$\alpha$ light curve of the superflare (red) and its background level (black). The background model derived from the data on March 14 is consistent with the March 11.
(E) The light curves of the superflare observed in H$\alpha$ (red) and \textit{TESS} white light (black). 
The lack of H$\alpha$ data at approximately  BJD-2458923 + 4 hrs is due to the temporary cloud in the sky.
The dashed line is the assumed background level for each.
}
\label{fig:2}
\end{figure}

Figure \ref{fig:2} (A) show the \textit{TESS} white-light's global light curves for a rotation phase of EK Dra around March 14, 2020. The period is derived as 2.62 days with Lomb-Scargle Periodogram (\textsf{astropy.stats.LombScargle}).
The TESS light curve shows quasi-periodic brightness variations over the whole observational period, indicating a gigantic starspot on the star \citep{2019ApJ...871..187N,2020ApJ...891..103N}.
As in the detrended light curve (Figure \ref{fig:2} (B)), a white-light flare is detected by \textit{TESS} on BJD-2458923 (March 14, 2020).
The amplitude is approximately 0.3 \%, which is significantly larger than the \textit{TESS} photometric errors (0.023 \%).

Simultaneously, we detected a clear H$\alpha$ flare (Figure \ref{fig:2} (C)). 
There is a rotational brightness variation in H$\alpha$, anti-correlated with the white light. 
We consider that the H$\alpha$ modulation is due to the background stellar active region and then subtracted the background trend to extract the flare light curve.
As in Figure \ref{fig:2} (D), we have modeled the background by linearly fitting the pre-flare and post-flare levels of March 14 (--0.4$\sim$0 hr and 6$\sim$7 hr), and obtained the flare light curve (Figure \ref{fig:2} (E)).

The white-light flare energy ($E_{\rm WL}$) is (2.6$\pm$0.3$)\times$10$^{34}$ erg, and the H$\alpha$ energy ($E_{\rm H\alpha}$) is (4.0$\pm$0.4)$\times$10$^{32}$ erg (1.5 \% of the white-light energy); thus, it is classified to a superflare.
This superflare is the 9th largest event in the six \textit{TESS} sectors (Figure \ref{fig:1} (E)).
As in Figure \ref{fig:2} (E), we found that the brightness evolution and timescales of the TESS white-light flare are almost the same as that of H$\alpha$ flare.
The full-width-at-half-maximum (FWHM) of the white-light flare ($t_{\rm WL}$) of and H$\alpha$ flare ($E_{\rm H\alpha}$) is 2.2 hrs and 2.3 hrs, respectively.

\begin{figure}
\epsscale{0.75}
\plotone{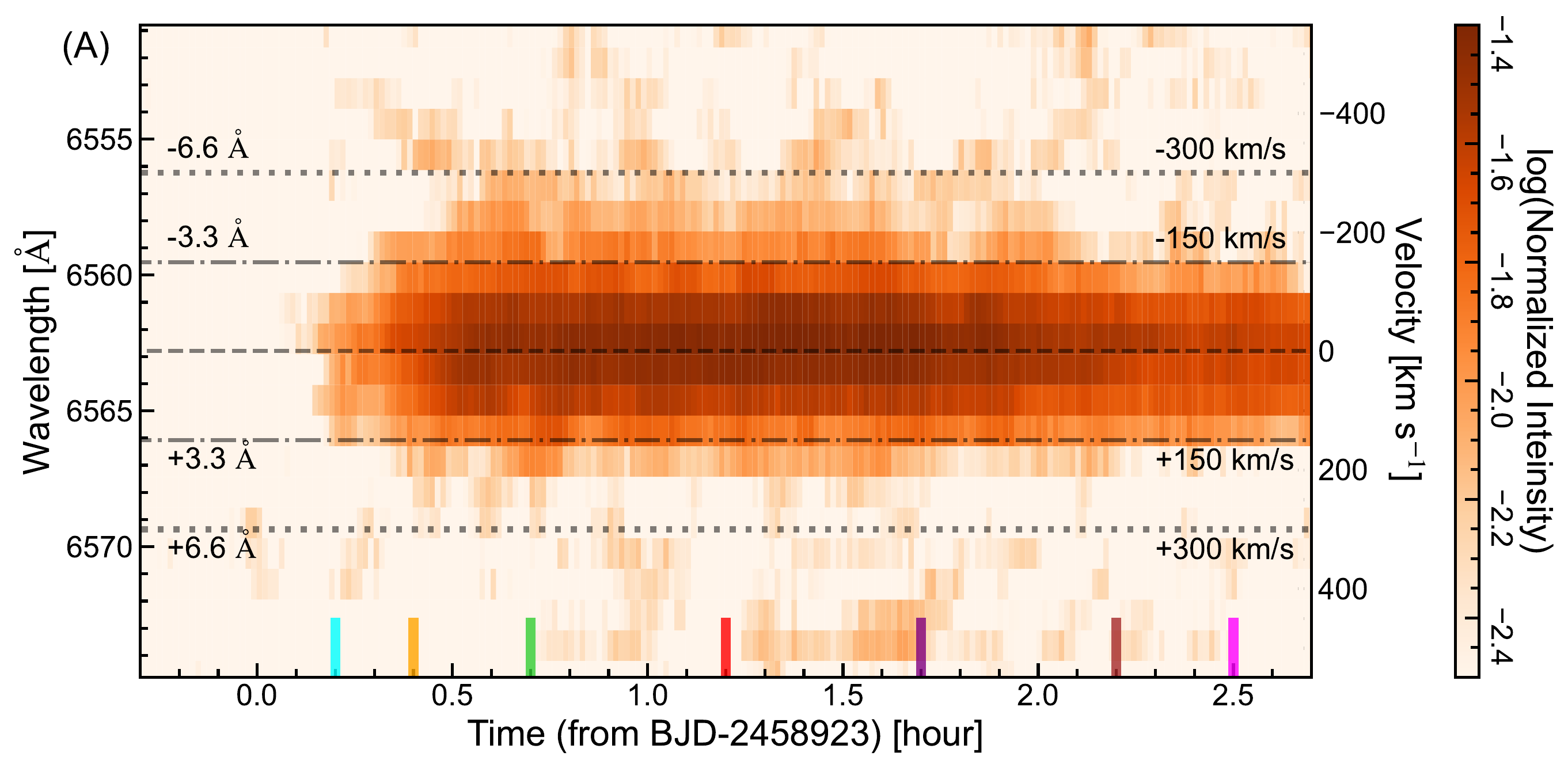}
\epsscale{1}
\plotone{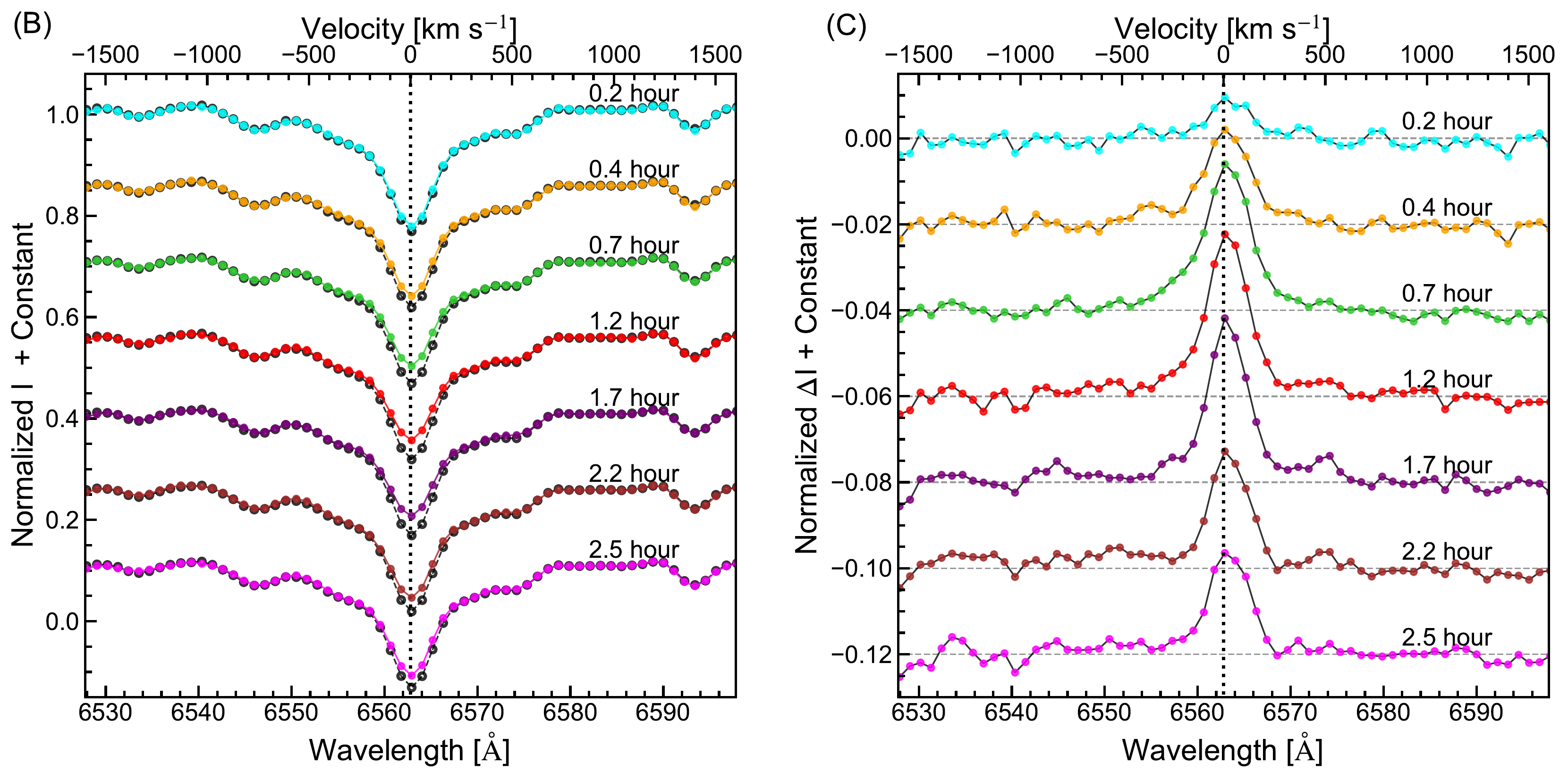}
\caption{The temporal evolution of H$\alpha$ line profiles of the superflare on EK Dra. 
(A) Temporal variation of the pre-flare-subtracted H$\alpha$ spectra in the time-wavelength plane. 
The color bar shows the emission intensity normalized by the continuum level (6517.8-6537.8 {\AA} and 6587.8-6607.8 {\AA}). 
(B) The H$\alpha$ spectra of the superflare on March 14, 2020 observed with Seimei Telescope/KOOLS-IFU. Each colored spectrum indicates the 20-min-averaged spectrum during the superflare with the central time indicated in the panel, and the background black dashed line is the pre-flare template spectrum. The pre-flare template spectrum was created by averaging the first 35 pieces of data for the night (-0.35 hr to 0.15 hr in panel (A)). The spectra are normalized by the continuum level and the constant values are added for visibility.
The dotted vertical line indicates the line center.
(C) The pre-flare subtracted spectra of those in the panel (B).
The basal level of each spectrum is plotted with the horizontal dashed line.
}
\label{fig:3}
\end{figure}
\begin{figure}
\epsscale{1}
\plottwo{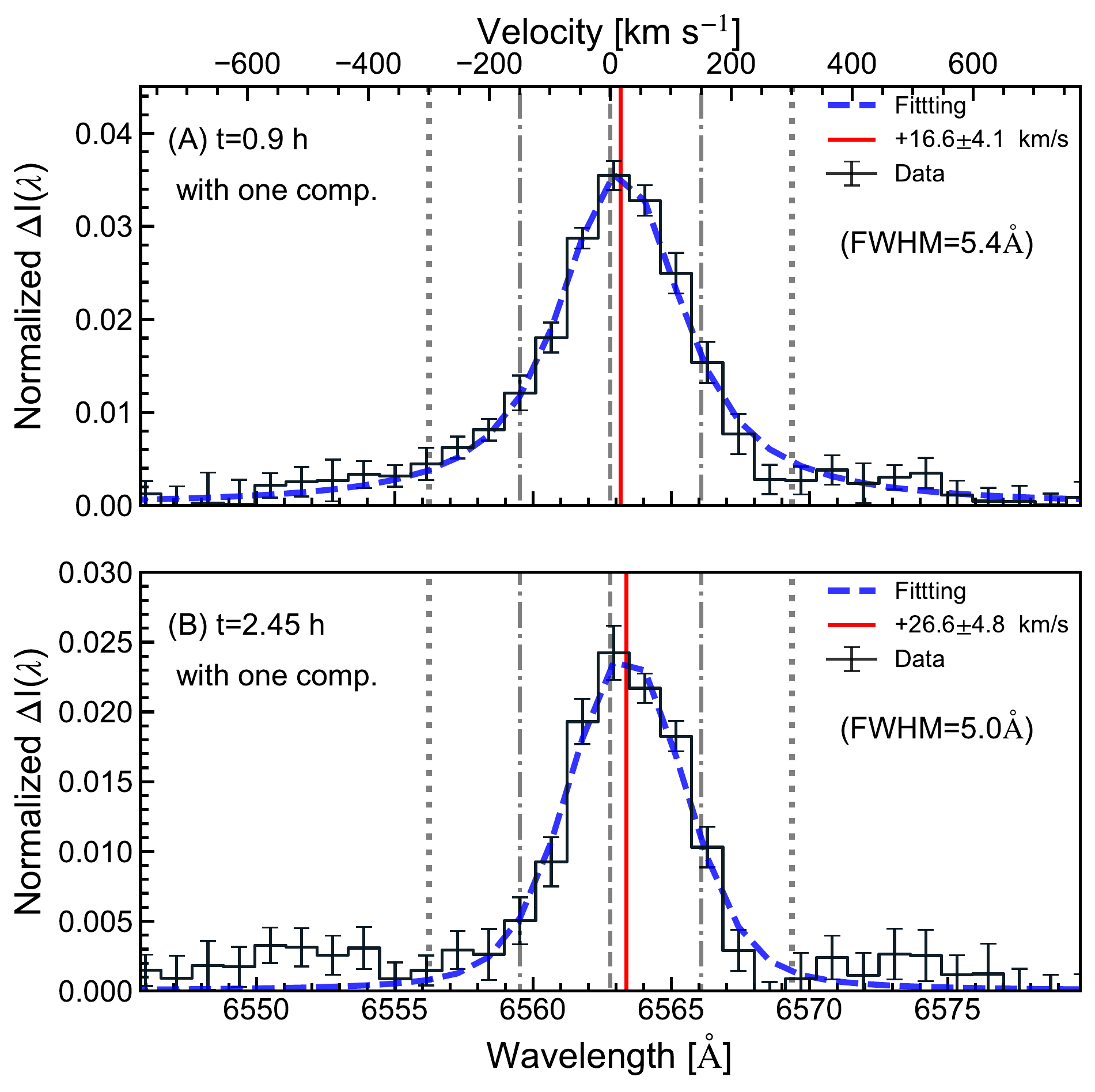}{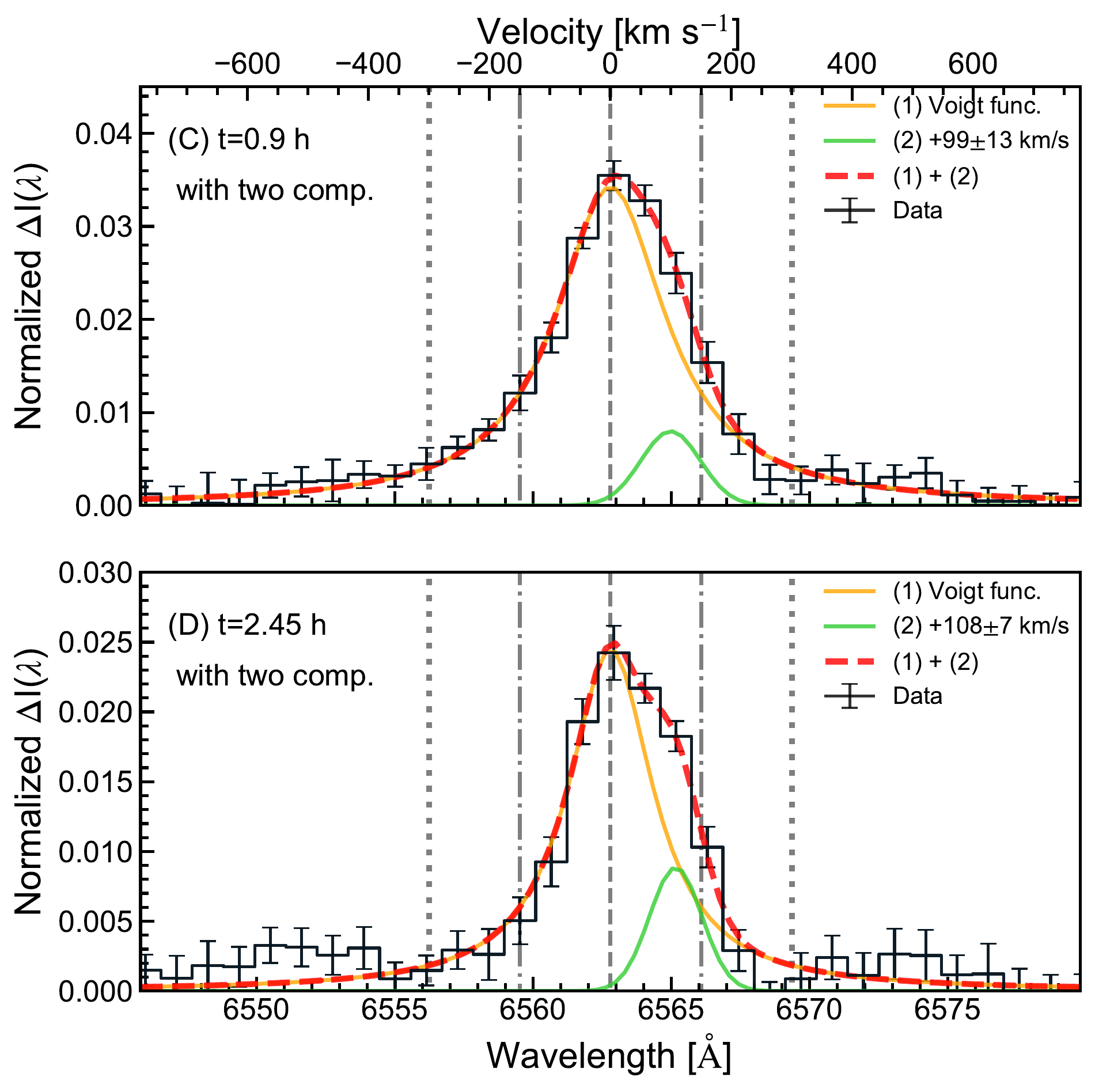}
\caption{The H$\alpha$ line profiles of the superflare on EK Dra. 
(A, B) The pre-flare-subtracted H$\alpha$ spectra at t=0.9 and 2.45 minutes, respectively. The blue lines result from the one component fitting with Voigt function. The red-vertical lines correspond to the fitted line center.
The line centers are 16.7$\pm$4.1 km s$^{-1}$ and 26.6$\pm$4.8 km s$^{-1}$, as indicated in each panel.
The FWHM of the lines are 5.4 {\AA} and 5.0 {\AA}, as also indicated in the panels.
(C, D) The black data are the same as the panels (A) and (B), respectively.
The colored line in the panels (C) and (D) are the results of two-component fitting with Voigt function (orange line for the central component) and Gaussian function (green line for the red-shift component).
We first fitted the blue-wing profile with Voigt function, and fitted the residual of the red wing with the Gauss function.
The red lines is the sum of green and orange lines.
The central velocity of the Gaussian (green lines) components of the panels (C) and (D) are 99$\pm$13 km s$^{-1}$ and 108$\pm$7 km s$^{-1}$, respectively.
The other properties of the red-wing enhancement are summarized in Table \ref{tab:1}.
}
\label{fig:4}
\end{figure}

Figure \ref{fig:3} (A-C) shows the time evolution of the pre-flare subtracted H$\alpha$ spectra.
We could not find any significant asymmetry of H$\alpha$ line profiles and line broadening having higher velocity than the instrumental resolution of 150 km s$^{-1}$.
Carefully looking at the spectra in Figure \ref{fig:4} (A-D), however, there may be a possible red asymmetry.
The redshift velocity is approximately 20 km s$^{-1}$ with one-component fitting (Figure \ref{fig:3} (A-B)), while it is approximately 100 km s$^{-1}$ with two-component fitting (Figure \ref{fig:4} (C-D)).
However, these are less than the instrumental dispersion velocity, and we would call this a ``possible redshift" and limit ourselves to speculating its possibilities in this Letter.

\section{Discussion} \label{sec:dis}

\begin{deluxetable*}{cccccccccd}
\tablenum{1}
\tablecaption{\Add{Summary of properties of the superflares on EK Dra and solar flares.\label{tab:1}}}
\tablewidth{0pt}
\tablehead{
\colhead{} & \colhead{$E_{\rm WL}$} & \colhead{$t_{\rm WL}$} & \colhead{$E_{\rm H\alpha}$/$E_{\rm WL}$} & \colhead{$t_{\rm H\alpha}/t_{\rm WL}$$^\S$} & \colhead{Asym.$^\dagger$} & \colhead{$I_{\rm red}$/$I_{\rm cen}$$^{\#}$} & \colhead{$V_{\rm redshift}$} & ref.  \\
\colhead{} & \colhead{[erg]} & & & & & &  \colhead{[km/s]}  
}
\startdata
EK Dra (Mar 14) & 2.6$\times$10$^{34}$ & 2.2 hr$^\S$ & 0.015 & 1.1 & No (Red) & ($\sim$0.36) & ($\sim$26.6$^{\ast}$/116$^{\#}$) & This Letter \\
\hline
EK Dra (Apr 5)$^{\ss}$ & 2.0$\times$10$^{33}$ & 16 min$^\%$ & (0.0085) & (1.1) & -- & -- & -- & ref.$^{(1)}$ \\
The Sun & 10$^{29-32}$$^{(2,3)}$ & 1-10 min$^{\%}$$^{(3)}$ & -- & $t_{\rm H\alpha}>$ $t_{\rm WL}$$^{(4)}$ & Red$^{\ae}$$^{(5,6)}$ & $\sim$several$\times$0.1$^{(5)}$ & $\le$150$^{(5)}$ & ref.$^{(2-6)}$  \\
\enddata
\tablecomments{\Add{``--" means no value and values in parentheses ``()"  are only reference values.
$^\S$ The FWHM duration of the flare. 
$^{\%}$The total duration of the flare.
$^\dagger$``Red" mean the red wind enhancement of the H$\alpha$ line emission profiles.
$^{\ast}$Single component fitting. $^{\#}$Two component fitting. 
$^{\ss}$The flare emission on April 5 in \cite{Namekata2020Sci} is very short-lived and its light curve is a combination of blueshift absorption and flaring emission. The H$\alpha$ flare duration is expected to be underestimated, and the line asymmetry of the H$\alpha$ flaring component was difficult to identify. Note that the  blueshifted ``absorption" in the flare on April 5 is not a flare radiation component, so the asymmetry is described here as ``--".}
$^{\ae}$Ref.$^{5}$ reported that only 5\% of solar flare show blue asymmetry and it is rare.
ref.$^{(1)}$: \cite{Namekata2020Sci}. ref.$^{(2)}$: \cite{2011LRSP....8....6S}. ref.$^{(3)}$: \cite{2017ApJ...851...91N}. ref.$^{(4)}$: \cite{2017NatCo...8.2202H}. ref.$^{(5)}$: \cite{1984SoPh...93..105I}. ref.$^{(6)}$: \cite{1962BAICz..13...37S}
}
\end{deluxetable*}


\subsection{The superflare on March 14, 2020 in comparison with solar and M-dwarf flares} \label{sec:4-1}

This section compares the superflare on March 14 with typical solar flares and M-dwarf (super-)flares.
One of the significant difference between typical solar flares and the superflare on EK Dra is the values of energy and duration \Add{(see Table \ref{tab:1})}.
\Add{The duration $\sim$2.2 hrs of the superflare with the energy of 2.6$\times10^{34}$ erg is more than ten times longer than those of solar white-light flares \citep[1-10 min for $10^{29-31}$ erg;][]{2017ApJ...851...91N}.}
The magnitude of these physical quantities is though to attribute to the flare length scale.
\cite{2017ApJ...851...91N} proposed that the length scale ($L$) \Add{of solar and stellar flaring loops} can be estimated from the \Add{white-light} flare energy ($E_{\rm flare}$) and e-folding decay time of \Add{white-light flare} ($\tau_{\rm decay}$ of 26 min) based on the magnetic reconnection theory in the formula of 
\begin{eqnarray}\label{eq:1}
L\sim 1.64\times 10^{9} \left( \frac{\tau_{\rm decay}}{100 \rm [s]} \right)^{2/5} \left( \frac{E_{\rm flare}}{10^{30} \rm [erg]} \right)^{1/5} \rm [cm].
\end{eqnarray}
This predicts the length scale as 3.8$\times 10^{10}$cm ($\sim$0.58 $R_{\rm star}$, $R_{\rm star}$ is stellar radius), which is much larger than the typical length scales of solar flares ($10^{9}\sim10^{10}$ cm).

Furthermore, the most mysterious point is that the duration of H$\alpha$ and white light is approximately the same for the EK Dra superflare.
In the case of solar flares \citep[e.g.,][\Add{and see Table \ref{tab:1}}]{2017NatCo...8.2202H} and M-dwarf flares \citep[e.g.,][]{1991ApJ...378..725H,2013ApJS..207...15K,2020PASJ...72...68N}, the white light timescale is generally several times shorter than that of H$\alpha$. 
For example, the ratio of H$\alpha$ and white-light duration was reported as a factor of 2 in the 10$^{33}$ erg-class M-dwarf superflares by \cite{2020PASJ...72...68N}.
Much larger difference was reported for several M-dwarf flares by \cite{2013ApJS..207...15K}.
Possible mechanisms for this unexpected results of the EK Dra superflare will be discussed in Section \ref{sec:4-4}.

In addition, the superflare did not show any significant line broadening synchronized (Figure \ref{fig:4}), which is often seen in M-dwarf superflares \citep[e.g., $\sim$14 {\AA} for a 10$^{33}$ erg-class superflare in][]{2020PASJ...72...68N} with the same spectrograph KOOLS-IFU. This indicates that high energy electrons heating is not essential for this superflare on EK Dra.

Other properties of stellar superflares can be explained by the analogy of solar and M-dwarf flares.
The energy of H$\alpha$ relative to the energy of white light is roughly only a few percent for both EK Dra superflare (Table \ref{tab:1}), and M-dwarf flares \citep[e.g.,][]{2020PASJ...72...68N}.
In addition, the upper limits of the estimated values of the ``possible" line asymmetry of H$\alpha$ line of this superflare do not contradict with typical values of solar flares \citep[\Add{see Table \ref{tab:1} and}][]{1984SoPh...93..105I} even though they are real.


\subsection{The superflare on March 14, 2020 in all other flares on EK Dra} \label{sec:4-2}

In Figure \ref{fig:1} (F), the relationship between flare energy and duration on EK Dra is compared with that on \textit{Kepler} solar-type stars.
According to \cite{2017ApJ...851...91N}, the relationship for superflares on \textit{Kepler} solar-type stars can be explained by the magnetic reconnection model (cf., Equation \ref{eq:1}).
Based on their theory, the consistency between EK Dra superflares and \textit{Kepler} superflares means that the magnetic reconnection model can also be applied to the EK Dra superflares.
Also, the superflare on March 14 is consistent with the majority of the flares on EK Dra, and this indicates that it was not a special case among other flares. 
 

In Table \ref{tab:1}, the superflare on EK Dra on March 14, 2020 is compared with that on April 5, 2020 reported by \cite{Namekata2020Sci}.
The duration and white-light energy of superflare on March 14, 2020  is ten times larger than those of superflare on April 5, 2020.
The significant difference between March 14 and April 5 events is the association of filament eruption signature \Add{(i.e., a blue-shifted H$\alpha$ absorption as in \citeauthor{Namekata2020Sci} \citeyear{Namekata2020Sci})}.
\Add{If a filament is in the front of the star and has a line-of-sight velocity, a blueshifted absorption should be observed \citep[e.g.,][and references therein]{Namekata2020Sci}. In other words, this result of March 14 event means that a filament eruption could occur either outside the stellar disk or perpendicular to the line of sight, meaning the flare could occur around the stellar limb (see the related discussion in Section \ref{sec:4-3} and \ref{sec:4-4}).
Other than these possibility, the lack of the signature of filament eruption on March 14 may indicate that mass ejection events do not always happen for every superflares, as not all flares are accompanied by CMEs in the case of the Sun \citep[e.g.,][]{2009IAUS..257..233Y}.}

\subsection{Relation between the superflare and spot groups} \label{sec:4-3}

The anti-correlated rotational variation in H$\alpha$ and white-light (Figure \ref{fig:2} (A, C)) is the same as the Sun-as-a-star active region modulation \Add{\citep[e.g.,][]{2019A&A...627A.118M}}, indicating that the chromosphere around the starspot (group) is magnetically heated and appears bright.
The spot area on EK Dra estimated from the white-light brightness variation is 0.032 of the stellar disk by using the method of \cite{2017PASJ...69...41M}, and the length scale is 0.18 of $R_{\rm star}$ (1.2$\times 10^{10}$ cm, in the photosphere), which is one third of the flare loop length scale of 0.58 $R_{\rm star}$ (in the corona).
The stored magnetic energy is roughly 1.2$\times 10^{36}$ erg when the mean magnetic field is 1000 G \Add{(see Equation 1 in \citeauthor{2013PASJ...65...49S} \citeyear{2013PASJ...65...49S}),}  meaning that the star had an energy enough to produce the superflares.


The rotational modulation is not entirely symmetric in time (Figure \ref{fig:2} (B)), suggesting the existence of multiple starspots (groups) rather than a single concentrated starspot.
This superflare occurs at +0.11 rotation period relative to the local maximum of the TESS light curve and at --0.19 rotation period relative to the local minimum. 
This means that the flare occurred when a giant spot (group) started to be visible by the stellar rotation.
If we assume that the superflare occur around the dominant spot, the superflare can have occurred near the stellar limb.
This may explain no signature of a filament eruption \Add{(Section \ref{sec:4-2})}.


\subsection{Possible emission mechanisms} \label{sec:4-4}

The most notable point of this superflare is that the duration and light curve evolution of H$\alpha$ are the same as those of white light as in Section \ref{sec:4-2}. 
In the standard model of solar flares, white-light flares are emitted from the photosphere or/and chromosphere heated by non-thermal electrons, while H$\alpha$ is emitted mainly from both the non-thermally and thermally heated chromosphere \citep[e.g.,][]{2011LRSP....8....6S,2016ApJ...820...95K}.
Since the timescale of thermal heating is, in general, longer than that of non-thermal heating, H$\alpha$ flares are thought to be qualitatively longer than white-light flares \citep[\Add{see Table \ref{tab:1}}, so-called \textit{Neupert} effect;][]{1968ApJ...153L..59N}. 
This standard model cannot explain the time evolution of the superflare on EK Dra.
In addition, the lack of H$\alpha$ line broadening may indicate that non-thermal heating at the footpoints, a main flare driver in large solar flares \citep{2017ApJ...836...17A}, is less dominant.

Then, what emission mechanism can explain the unexpected time evolution of the superflare?
We propose the following two hypotheses: 
First, \cite{2019MNRAS.489.4338N} proposed that the X-ray back-warming can produce white-light enhancement in the photosphere in such an extreme case.
\Add{Although the radiative back-warming is proposed as one contributor to white-light emission even for solar flares \citep[e.g.,][]{2017NatCo...8.2202H}, how dominant it is is unknown.}
If this process can be applied, both white-light and H$\alpha$ emissions can last as long as the thermal emission in the coronal loop does.
If the heated photosphere covers 10\% of the stellar disk ($\sim$ the square of the length scale), the white-light enhancement becomes $\sim$0.8\% of $L_{\rm star}$ \Add{according to the numerical modeling by \cite{2019MNRAS.489.4338N}}. 

Second, \cite{2018ApJ...859..143H} theoretically proposed that the contribution of flare loops to not only H$\alpha$ but also white-light flares can be large in the case of superflares.
If this mechanism is possible, the duration and evolution could be the same since both H$\alpha$ and white-light radiation have the same emission source (i.e., flare loops).
\Add{As in Section \ref{sec:4-3} and \ref{sec:4-4}}, if the flare occurred around the limb (i.e. an ``off-limb" flare), the flare loop emission can be more dominant.
Because the area of flare loop is $\sim$10\% of the disk area for this superflare, the flare loop can produce the observed white-light enhancement ($\sim$0.02 of stellar luminosity) according to \cite{2018ApJ...859..143H}, if the flare-loop density is sufficiently high ($\sim$10$^{13}$ cm$^{-3}$).
The H$\alpha$ emission of the flare loop in the superflare case is expected to be approximately 1-5\% of the continuum level \citep[cf.][]{1996SoPh..166...89W}, which also corresponds well with observations.

The above two models are not supposed to be dominant in solar flares.
However, our observation may indicate that they may present \Add{radiation mechanisms} unique to the giant superflare. 

\subsection{Implications and future works} \label{sec:5}



Section \ref{sec:4-4} suggest a possible radiation mechanism different from that of general solar flares. However, it is not clear from this one example alone how universal it is. 
More samples are needed, and the analysis of solar flares in the Sun-as-a-star view \citep{Namekata2020Sci} and the numerical modelings of flares \citep[e.g.,][]{2020PASJ...72...68N} will help interpretations.
In addition, our superflare did not show a signature of a filament eruption, unlike the only previous detection of a superflare on a solar-type star \citep{Namekata2020Sci}. 
\Add{It is known that in the case of the Sun, not all flares are accompanied by CMEs \citep{2009IAUS..257..233Y}, but the CME-association rates for stellar superflares are unknown.}
Further observations may help us estimate the frequency of filament eruptions/CMEs, which is essential information for the planetary habitability around the young Sun \citep[cf.][]{2020IJAsB..19..136A} and for the estimation of a CME-related mass-loss rates \citep[e.g.,][]{2015ApJ...809...79O}. 


\Add{Finally, the superflare on the solar-type star observed in this Letter is important as a proxy for a possible superflare that may occur on the present-day Sun.
Interestingly, the energy of the observed superflare is comparable to the estimated upper limit of flare energy on old Sun-like stars and the present-day Sun \citep[$\sim$4$ \times $10$^{34}$ erg;][]{2020arXiv201102117O}, so the revealed properties in this Letter could be helpful to model the chromospheric radiations from a possible extreme superflare on the Sun.}

\begin{acknowledgments}
The spectroscopic data presented here were obtained at the Okayama Observatory of Kyoto University, which is operated as a scientific partnership with National Astronomical Observatory of Japan. 
This work and operations of OISTER were supported by the Optical and Near-infrared Astronomy Inter-University Cooperation Program and the Grants-in-Aid of the Ministry of Education. 
Funding for the TESS mission is provided by NASA's Science Mission directorate.
We would like to thank S. Toriumi for discussion and comments.
We acknowledges the International Space Science Institute and the supported International Team 464: The Role of Solar and Stellar Energetic Particles on (Exo) Planetary Habitability (ETERNAL, \url{http://www.issibern.ch/ teams/exoeternal/}).
Y.N. was supported by the JSPS Overseas Research Fellowship Program.
This research is supported by JSPS KAKENHI grant numbers 18J20048, 21J00316 (K.N.), 17K05400, 20K04032, 20H05643 (H.M.), 21J00106 (Y.N.), 20K14521 (K.I.), and 21H01131 (K. Shibata, H.M., S.H. and D.N.). 
\end{acknowledgments}

%

\vspace{5mm}


\software{\textsf{IRAF} \citep{Tody1986}, \textsf{PyRAF} \citep{2012ascl.soft07011S}}


\clearpage

\clearpage

\clearpage
\bibliography{sample631,cvs}{}

\newcommand{\noop}[1]{}
\begin{thebibliography}{}
\expandafter\ifx\csname natexlab\endcsname\relax\def\natexlab#1{#1}\fi
\providecommand{\url}[1]{\href{#1}{#1}}
\providecommand{\dodoi}[1]{doi:~\href{http://doi.org/#1}{\nolinkurl{#1}}}
\providecommand{\doeprint}[1]{\href{http://ascl.net/#1}{\nolinkurl{http://ascl.net/#1}}}
\providecommand{\doarXiv}[1]{\href{https://arxiv.org/abs/#1}{\nolinkurl{https://arxiv.org/abs/#1}}}

\bibitem[{{Airapetian} {et~al.}(2020){Airapetian}, {Barnes}, {Cohen},
  {Collinson}, {Danchi}, {Dong}, {Del Genio}, {France}, {Garcia-Sage},
  {Glocer}, {Gopalswamy}, {Grenfell}, {Gronoff}, {G{\"u}del}, {Herbst},
  {Henning}, {Jackman}, {Jin}, {Johnstone}, {Kaltenegger}, {Kay}, {Kobayashi},
  {Kuang}, {Li}, {Lynch}, {L{\"u}ftinger}, {Luhmann}, {Maehara}, {Mlynczak},
  {Notsu}, {Osten}, {Ramirez}, {Rugheimer}, {Scheucher}, {Schlieder},
  {Shibata}, {Sousa-Silva}, {Stamenkovi{\'c}}, {Strangeway}, {Usmanov},
  {Vergados}, {Verkhoglyadova}, {Vidotto}, {Voytek}, {Way}, {Zank}, \&
  {Yamashiki}}]{2020IJAsB..19..136A}
{Airapetian}, V.~S., {Barnes}, R., {Cohen}, O., {et~al.} 2020, International
  Journal of Astrobiology, 19, 136, \dodoi{10.1017/S1473550419000132}

\bibitem[{{Aschwanden} {et~al.}(2017){Aschwanden}, {Caspi}, {Cohen}, {Holman},
  {Jing}, {Kretzschmar}, {Kontar}, {McTiernan}, {Mewaldt}, {O'Flannagain},
  {Richardson}, {Ryan}, {Warren}, \& {Xu}}]{2017ApJ...836...17A}
{Aschwanden}, M.~J., {Caspi}, A., {Cohen}, C. M.~S., {et~al.} 2017, \apj, 836,
  17, \dodoi{10.3847/1538-4357/836/1/17}

\bibitem[{{Aschwanden}(2019)}]{2019ApJ...880..105A}
{Aschwanden}, M.~J. 2019, \apj, 880, 105, \dodoi{10.3847/1538-4357/ab29f4}

\bibitem[{{Audard} {et~al.}(1999){Audard}, {G{\"u}del}, \&
  {Guinan}}]{1999ApJ...513L..53A}
{Audard}, M., {G{\"u}del}, M., \& {Guinan}, E.~F. 1999, \apjl, 513, L53,
  \dodoi{10.1086/311907}

\bibitem[{{Aulanier} {et~al.}(2013){Aulanier}, {D{\'e}moulin}, {Schrijver},
  {Janvier}, {Pariat}, \& {Schmieder}}]{2013A&A...549A..66A}
{Aulanier}, G., {D{\'e}moulin}, P., {Schrijver}, C.~J., {et~al.} 2013, \aap,
  549, A66, \dodoi{10.1051/0004-6361/201220406}

\bibitem[{{Ayres}(2015)}]{2015AJ....150....7A}
{Ayres}, T.~R. 2015, \aj, 150, 7, \dodoi{10.1088/0004-6256/150/1/7}

\bibitem[{{Benz}(2017)}]{2017LRSP...14....2B}
{Benz}, A.~O. 2017, Living Reviews in Solar Physics, 14, 2,
  \dodoi{10.1007/s41116-016-0004-3}

\bibitem[{{{\c{S}}enavc{\i}} {et~al.}(2021){{\c{S}}enavc{\i}},
  {K{\i}l{\i}{\c{c}}o{\u{g}}lu}, {I{\c{s}}{\i}k}, {Hussain}, {Montes}, {Bahar},
  \& {Solanki}}]{2021MNRAS.502.3343S}
{{\c{S}}enavc{\i}}, H.~V., {K{\i}l{\i}{\c{c}}o{\u{g}}lu}, T., {I{\c{s}}{\i}k},
  E., {et~al.} 2021, \mnras, 502, 3343, \dodoi{10.1093/mnras/stab199}

\bibitem[{{Emslie} {et~al.}(2012){Emslie}, {Dennis}, {Shih}, {Chamberlin},
  {Mewaldt}, {Moore}, {Share}, {Vourlidas}, \& {Welsch}}]{2012ApJ...759...71E}
{Emslie}, A.~G., {Dennis}, B.~R., {Shih}, A.~Y., {et~al.} 2012, \apj, 759, 71,
  \dodoi{10.1088/0004-637X/759/1/71}

\bibitem[{{Fausnaugh}(2020)}]{Fausnaugh2020}
{Fausnaugh}, M. M. e.~a. 2020, TESS Data Release Notes: Sector 23, DR32

\bibitem[{{Guarcello} {et~al.}(2019){Guarcello}, {Micela}, {Sciortino},
  {L{\'o}pez-Santiago}, {Argiroffi}, {Reale}, {Flaccomio},
  {Alvarado-G{\'o}mez}, {Antoniou}, {Drake}, {Pillitteri}, {Rebull}, \&
  {Stauffer}}]{2019A&A...622A.210G}
{Guarcello}, M.~G., {Micela}, G., {Sciortino}, S., {et~al.} 2019, \aap, 622,
  A210, \dodoi{10.1051/0004-6361/201834370}

\bibitem[{{Hao} {et~al.}(2017){Hao}, {Yang}, {Cheng}, {Guo}, {Fang}, {Ding},
  {Chen}, \& {Li}}]{2017NatCo...8.2202H}
{Hao}, Q., {Yang}, K., {Cheng}, X., {et~al.} 2017, Nature Communications, 8,
  2202, \dodoi{10.1038/s41467-017-02343-0}

\bibitem[{{Hawley} \& {Pettersen}(1991)}]{1991ApJ...378..725H}
{Hawley}, S.~L., \& {Pettersen}, B.~R. 1991, \apj, 378, 725,
  \dodoi{10.1086/170474}

\bibitem[{{Hayakawa} {et~al.}(2017){Hayakawa}, {Iwahashi}, {Ebihara},
  {Tamazawa}, {Shibata}, {Knipp}, {Kawamura}, {Hattori}, {Mase}, {Nakanishi},
  \& {Isobe}}]{2017ApJ...850L..31H}
{Hayakawa}, H., {Iwahashi}, K., {Ebihara}, Y., {et~al.} 2017, \apjl, 850, L31,
  \dodoi{10.3847/2041-8213/aa9661}

\bibitem[{{Heinzel} \& {Shibata}(2018)}]{2018ApJ...859..143H}
{Heinzel}, P., \& {Shibata}, K. 2018, \apj, 859, 143,
  \dodoi{10.3847/1538-4357/aabe78}

\bibitem[{{Ichimoto} \& {Kurokawa}(1984)}]{1984SoPh...93..105I}
{Ichimoto}, K., \& {Kurokawa}, H. 1984, \solphys, 93, 105,
  \dodoi{10.1007/BF00156656}

\bibitem[{{J{\"a}rvinen} {et~al.}(2018){J{\"a}rvinen}, {Strassmeier},
  {Carroll}, {Ilyin}, \& {Weber}}]{2018A&A...620A.162J}
{J{\"a}rvinen}, S.~P., {Strassmeier}, K.~G., {Carroll}, T.~A., {Ilyin}, I., \&
  {Weber}, M. 2018, \aap, 620, A162, \dodoi{10.1051/0004-6361/201833496}

\bibitem[{{Kowalski} {et~al.}(2013){Kowalski}, {Hawley}, {Wisniewski}, {Osten},
  {Hilton}, {Holtzman}, {Schmidt}, \& {Davenport}}]{2013ApJS..207...15K}
{Kowalski}, A.~F., {Hawley}, S.~L., {Wisniewski}, J.~P., {et~al.} 2013, \apjs,
  207, 15, \dodoi{10.1088/0067-0049/207/1/15}

\bibitem[{{Kowalski} {et~al.}(2016){Kowalski}, {Mathioudakis}, {Hawley},
  {Wisniewski}, {Dhillon}, {Marsh}, {Hilton}, \& {Brown}}]{2016ApJ...820...95K}
{Kowalski}, A.~F., {Mathioudakis}, M., {Hawley}, S.~L., {et~al.} 2016, \apj,
  820, 95, \dodoi{10.3847/0004-637X/820/2/95}

\bibitem[{{Kretzschmar}(2011)}]{2011A&A...530A..84K}
{Kretzschmar}, M. 2011, \aap, 530, A84, \dodoi{10.1051/0004-6361/201015930}

\bibitem[{{Kurita} {et~al.}(2020){Kurita}, {Kino}, {Iwamuro}, {Ohta}, {Nogami},
  {Izumiura}, {Yoshida}, {Matsubayashi}, {Kuroda}, {Nakatani}, {Yamamoto},
  {Tsutsui}, {Iribe}, {Jikuya}, {Ohtani}, {Shibata}, {Takahashi}, {Tokoro},
  {Maihara}, \& {Nagata}}]{2020PASJ...72...48K}
{Kurita}, M., {Kino}, M., {Iwamuro}, F., {et~al.} 2020, \pasj, 72, 48,
  \dodoi{10.1093/pasj/psaa036}

\bibitem[{{Lindegren} {et~al.}(2018){Lindegren}, {Hern{\'a}ndez}, {Bombrun},
  {Klioner}, {Bastian}, {Ramos-Lerate}, {de Torres}, {Steidelm{\"u}ller},
  {Stephenson}, {Hobbs}, {Lammers}, {Biermann}, {Geyer}, {Hilger}, {Michalik},
  {Stampa}, {McMillan}, {Casta{\~n}eda}, {Clotet}, {Comoretto}, {Davidson},
  {Fabricius}, {Gracia}, {Hambly}, {Hutton}, {Mora}, {Portell}, {van Leeuwen},
  {Abbas}, {Abreu}, {Altmann}, {Andrei}, {Anglada}, {Balaguer-N{\'u}{\~n}ez},
  {Barache}, {Becciani}, {Bertone}, {Bianchi}, {Bouquillon}, {Bourda},
  {Br{\"u}semeister}, {Bucciarelli}, {Busonero}, {Buzzi}, {Cancelliere},
  {Carlucci}, {Charlot}, {Cheek}, {Crosta}, {Crowley}, {de Bruijne}, {de
  Felice}, {Drimmel}, {Esquej}, {Fienga}, {Fraile}, {Gai}, {Garralda},
  {Gonz{\'a}lez-Vidal}, {Guerra}, {Hauser}, {Hofmann}, {Holl}, {Jordan},
  {Lattanzi}, {Lenhardt}, {Liao}, {Licata}, {Lister}, {L{\"o}ffler},
  {Marchant}, {Martin-Fleitas}, {Messineo}, {Mignard}, {Morbidelli}, {Poggio},
  {Riva}, {Rowell}, {Salguero}, {Sarasso}, {Sciacca}, {Siddiqui}, {Smart},
  {Spagna}, {Steele}, {Taris}, {Torra}, {van Elteren}, {van Reeven}, \&
  {Vecchiato}}]{2018A&A...616A...2L}
{Lindegren}, L., {Hern{\'a}ndez}, J., {Bombrun}, A., {et~al.} 2018, \aap, 616,
  A2, \dodoi{10.1051/0004-6361/201832727}

\bibitem[{{Maehara} {et~al.}(2012){Maehara}, {Shibayama}, {Notsu}, {Notsu},
  {Nagao}, {Kusaba}, {Honda}, {Nogami}, \& {Shibata}}]{2012Natur.485..478M}
{Maehara}, H., {Shibayama}, T., {Notsu}, S., {et~al.} 2012, Nature, 485, 478,
  \dodoi{10.1038/nature11063}

\bibitem[{{Maehara} {et~al.}(2015){Maehara}, {Shibayama}, {Notsu}, {Notsu},
  {Honda}, {Nogami}, \& {Shibata}}]{2015EP&S...67...59M}
{Maehara}, H., {Shibayama}, T., {Notsu}, Y., {et~al.} 2015, Earth, Planets, and
  Space, 67, 59, \dodoi{10.1186/s40623-015-0217-z}

\bibitem[{{Maehara} {et~al.}(2017){Maehara}, {Notsu}, {Notsu}, {Namekata},
  {Honda}, {Ishii}, {Nogami}, \& {Shibata}}]{2017PASJ...69...41M}
{Maehara}, H., {Notsu}, Y., {Notsu}, S., {et~al.} 2017, \pasj, 69, 41,
  \dodoi{10.1093/pasj/psx013}

\bibitem[{{Maehara} {et~al.}(2021){Maehara}, {Notsu}, {Namekata}, {Honda},
  {Kowalski}, {Katoh}, {Ohshima}, {Iida}, {Oeda}, {Murata}, {Yamanaka},
  {Takagi}, {Sasada}, {Akitaya}, {Ikuta}, {Okamoto}, {Nogami}, \&
  {Shibata}}]{2020PASJ..tmp..253M}
{Maehara}, H., {Notsu}, Y., {Namekata}, K., {et~al.} 2021, \pasj,
  \dodoi{10.1093/pasj/psaa098}

\bibitem[{{Maldonado} {et~al.}(2019){Maldonado}, {Phillips}, {Dumusque},
  {Collier Cameron}, {Haywood}, {Lanza}, {Micela}, {Mortier}, {Saar},
  {Sozzetti}, {Rice}, {Milbourne}, {Cecconi}, {Cegla}, {Cosentino}, {Costes},
  {Ghedina}, {Gonzalez}, {Guerra}, {Hern{\'a}ndez}, {Li}, {Lodi}, {Malavolta},
  {Molinari}, {Pepe}, {Piotto}, {Poretti}, {Sasselov}, {San Juan}, {Thompson},
  {Udry}, \& {Watson}}]{2019A&A...627A.118M}
{Maldonado}, J., {Phillips}, D.~F., {Dumusque}, X., {et~al.} 2019, \aap, 627,
  A118, \dodoi{10.1051/0004-6361/201935233}

\bibitem[{{Matsubayashi} {et~al.}(2019){Matsubayashi}, {Ohta}, {Iwamuro},
  {Iwata}, {Kambe}, {Tsutsui}, {Izumiura}, {Yoshida}, \&
  {Hattori}}]{2019PASJ...71..102M}
{Matsubayashi}, K., {Ohta}, K., {Iwamuro}, F., {et~al.} 2019, \pasj, 71, 102,
  \dodoi{10.1093/pasj/psz087}

\bibitem[{{Miyake} {et~al.}(2012){Miyake}, {Nagaya}, {Masuda}, \&
  {Nakamura}}]{2012Natur.486..240M}
{Miyake}, F., {Nagaya}, K., {Masuda}, K., \& {Nakamura}, T. 2012, \nat, 486,
  240, \dodoi{10.1038/nature11123}

\bibitem[{{Namekata} {et~al.}(2017){Namekata}, {Sakaue}, {Watanabe}, {Asai},
  {Maehara}, {Notsu}, {Notsu}, {Honda}, {Ishii}, {Ikuta}, {Nogami}, \&
  {Shibata}}]{2017ApJ...851...91N}
{Namekata}, K., {Sakaue}, T., {Watanabe}, K., {et~al.} 2017, \apj, 851, 91,
  \dodoi{10.3847/1538-4357/aa9b34}

\bibitem[{{Namekata} {et~al.}(2019){Namekata}, {Maehara}, {Notsu}, {Toriumi},
  {Hayakawa}, {Ikuta}, {Notsu}, {Honda}, {Nogami}, \&
  {Shibata}}]{2019ApJ...871..187N}
{Namekata}, K., {Maehara}, H., {Notsu}, Y., {et~al.} 2019, \apj, 871, 187,
  \dodoi{10.3847/1538-4357/aaf471}

\bibitem[{{Namekata} {et~al.}(2020{\natexlab{a}}){Namekata}, {Davenport},
  {Morris}, {Hawley}, {Maehara}, {Notsu}, {Toriumi}, {Ikuta}, {Notsu}, {Honda},
  {Nogami}, \& {Shibata}}]{2020ApJ...891..103N}
{Namekata}, K., {Davenport}, J. R.~A., {Morris}, B.~M., {et~al.}
  2020{\natexlab{a}}, \apj, 891, 103, \dodoi{10.3847/1538-4357/ab7384}

\bibitem[{{Namekata} {et~al.}(2020{\natexlab{b}}){Namekata}, {Maehara},
  {Sasaki}, {Kawai}, {Notsu}, {Kowalski}, {Allred}, {Iwakiri}, {Tsuboi},
  {Murata}, {Niwano}, {Shiraishi}, {Adachi}, {Iida}, {Oeda}, {Honda}, {Tozuka},
  {Katoh}, {Onozato}, {Okamoto}, {Isogai}, {Kimura}, {Kojiguchi}, {Wakamatsu},
  {Tampo}, {Nogami}, \& {Shibata}}]{2020PASJ...72...68N}
{Namekata}, K., {Maehara}, H., {Sasaki}, R., {et~al.} 2020{\natexlab{b}},
  \pasj, 72, 68, \dodoi{10.1093/pasj/psaa051}

\bibitem[{{Namekata} {et~al.}(2021){Namekata}, {Maehara}, {Honda}, {Notsu},
  {Okamoto}, {Takahashi}, {Takayama}, \& {Ohshima}}]{Namekata2020Sci}
{Namekata}, K., {Maehara}, H., {Honda}, S., {et~al.} 2021, Nature Astronomy,
  \dodoi{10.1038/s41550-021-01532-8}

\bibitem[{{Neupert}(1968)}]{1968ApJ...153L..59N}
{Neupert}, W.~M. 1968, \apjl, 153, L59, \dodoi{10.1086/180220}

\bibitem[{{Nizamov}(2019)}]{2019MNRAS.489.4338N}
{Nizamov}, B.~A. 2019, \mnras, 489, 4338, \dodoi{10.1093/mnras/stz2478}

\bibitem[{{Notsu} {et~al.}(2019){Notsu}, {Maehara}, {Honda}, {Hawley},
  {Davenport}, {Namekata}, {Notsu}, {Ikuta}, {Nogami}, \&
  {Shibata}}]{2019ApJ...876...58N}
{Notsu}, Y., {Maehara}, H., {Honda}, S., {et~al.} 2019, \apj, 876, 58,
  \dodoi{10.3847/1538-4357/ab14e6}

\bibitem[{{Okamoto} {et~al.}(2021){Okamoto}, {Notsu}, {Maehara}, {Namekata},
  {Honda}, {Ikuta}, {Nogami}, \& {Shibata}}]{2020arXiv201102117O}
{Okamoto}, S., {Notsu}, Y., {Maehara}, H., {et~al.} 2021, \apj, 906, 72,
  \dodoi{10.3847/1538-4357/abc8f5}

\bibitem[{{Osten} \& {Wolk}(2015)}]{2015ApJ...809...79O}
{Osten}, R.~A., \& {Wolk}, S.~J. 2015, \apj, 809, 79,
  \dodoi{10.1088/0004-637X/809/1/79}

\bibitem[{{Ricker} {et~al.}(2015){Ricker}, {Winn}, {Vanderspek}, {Latham},
  {Bakos}, {Bean}, {Berta-Thompson}, {Brown}, {Buchhave}, {Butler}, {Butler},
  {Chaplin}, {Charbonneau}, {Christensen-Dalsgaard}, {Clampin}, {Deming},
  {Doty}, {De Lee}, {Dressing}, {Dunham}, {Endl}, {Fressin}, {Ge}, {Henning},
  {Holman}, {Howard}, {Ida}, {Jenkins}, {Jernigan}, {Johnson}, {Kaltenegger},
  {Kawai}, {Kjeldsen}, {Laughlin}, {Levine}, {Lin}, {Lissauer}, {MacQueen},
  {Marcy}, {McCullough}, {Morton}, {Narita}, {Paegert}, {Palle}, {Pepe},
  {Pepper}, {Quirrenbach}, {Rinehart}, {Sasselov}, {Sato}, {Seager},
  {Sozzetti}, {Stassun}, {Sullivan}, {Szentgyorgyi}, {Torres}, {Udry}, \&
  {Villasenor}}]{2015JATIS...1a4003R}
{Ricker}, G.~R., {Winn}, J.~N., {Vanderspek}, R., {et~al.} 2015, Journal of
  Astronomical Telescopes, Instruments, and Systems, 1, 014003,
  \dodoi{10.1117/1.JATIS.1.1.014003}

\bibitem[{{Science Software Branch at STScI}(2012)}]{2012ascl.soft07011S}
{Science Software Branch at STScI}. 2012, {PyRAF: Python alternative for IRAF}.
\newblock \doeprint{1207.011}

\bibitem[{{Shibata} \& {Magara}(2011)}]{2011LRSP....8....6S}
{Shibata}, K., \& {Magara}, T. 2011, Living Reviews in Solar Physics, 8, 6,
  \dodoi{10.12942/lrsp-2011-6}

\bibitem[{{Shibata} {et~al.}(2013){Shibata}, {Isobe}, {Hillier}, {Choudhuri},
  {Maehara}, {Ishii}, {Shibayama}, {Notsu}, {Notsu}, {Nagao}, {Honda}, \&
  {Nogami}}]{2013PASJ...65...49S}
{Shibata}, K., {Isobe}, H., {Hillier}, A., {et~al.} 2013, \pasj, 65, 49,
  \dodoi{10.1093/pasj/65.3.49}

\bibitem[{{Shibayama} {et~al.}(2013){Shibayama}, {Maehara}, {Notsu}, {Notsu},
  {Nagao}, {Honda}, {Ishii}, {Nogami}, \& {Shibata}}]{2013ApJS..209....5S}
{Shibayama}, T., {Maehara}, H., {Notsu}, S., {et~al.} 2013, \apjs, 209, 5,
  \dodoi{10.1088/0067-0049/209/1/5}

\bibitem[{{Temmer}(2021)}]{2021LRSP...18....4T}
{Temmer}, M. 2021, Living Reviews in Solar Physics, 18, 4,
  \dodoi{10.1007/s41116-021-00030-3}

\bibitem[{{Tody}(1986)}]{Tody1986}
{Tody}, D. 1986, in Society of Photo-Optical Instrumentation Engineers (SPIE)
  Conference Series, Vol. 627, Instrumentation in astronomy VI, ed. D. L.
  Crawford, 733

\bibitem[{{{\v{S}}vestka} {et~al.}(1962){{\v{S}}vestka}, {Kopeck{\'y}}, \&
  {Blaha}}]{1962BAICz..13...37S}
{{\v{S}}vestka}, Z., {Kopeck{\'y}}, M., \& {Blaha}, M. 1962, Bulletin of the
  Astronomical Institutes of Czechoslovakia, 13, 37

\bibitem[{{Waite} {et~al.}(2017){Waite}, {Marsden}, {Carter}, {Petit},
  {Jeffers}, {Morin}, {Vidotto}, {Donati}, \& {BCool
  Collaboration}}]{2017MNRAS.465.2076W}
{Waite}, I.~A., {Marsden}, S.~C., {Carter}, B.~D., {et~al.} 2017, \mnras, 465,
  2076, \dodoi{10.1093/mnras/stw2731}

\bibitem[{{Wiik} {et~al.}(1996){Wiik}, {Schmieder}, {Heinzel}, \&
  {Roudier}}]{1996SoPh..166...89W}
{Wiik}, J.~E., {Schmieder}, B., {Heinzel}, P., \& {Roudier}, T. 1996, \solphys,
  166, 89, \dodoi{10.1007/BF00179357}

\bibitem[{{Yashiro} \& {Gopalswamy}(2009)}]{2009IAUS..257..233Y}
{Yashiro}, S., \& {Gopalswamy}, N. 2009, in Universal Heliophysical Processes,
  ed. N.~{Gopalswamy} \& D.~F. {Webb}, Vol. 257, 233--243,
  \dodoi{10.1017/S1743921309029342}

\end{thebibliography}
\bibliographystyle{aasjournal}



\end{document}